\documentclass[global,taphys]{svjour}%
\usepackage{amsfonts}
\usepackage{amsmath}
\usepackage{makeidx}
\usepackage{graphicx}
\usepackage{multicol}
\usepackage{amssymb}%
\begin{document}

\journalname{Adv. in Solid State Physics}
\title{From Digital to Analogue Magnetoelectronics: Theory of Transport in
Non-Collinear Magnetic Nanostructures}
\author{Gerrit E.W. Bauer,\inst{1} Yaroslav Tserkovnyak,\inst{2} Daniel
Huertas-Hernando\inst{1,3} and Arne Brataas\inst{4}}
\institute{Department of NanoScience, Delft University of Technology, Lorentzweg 1, 2628
CJ Delft, The Netherlands,
\email{G.E.W.Bauer@TNW.TUDelft.NL}%
\and Lyman Laboratory of Physics, Harvard University, Cambridge, Massachusetts
02138 USA \and Department of Physics, Sloane Physics Laboratory, Yale
University, New Haven, CT 06520-8120, USA\and Department of Physics, Norwegian
University of Science and Technology, N-7491.}
\titlerunning{Digital to analogue}%
\authorrunning{Gerrit Bauer et al.}%
\maketitle

\begin{abstract}
Magnetoelectronics is mainly digital, \textit{i.e.} governed by up and down
magnetizations. In contrast, analogue magnetoelectronics makes use of
phenomena occuring for non-collinear magnetization configurations. Here we
review theories which have recently been applied to the transport in
non-collinear magnetic nanostructures in two and multiterminal structures,
\textit{viz}. random matrix and circuit theory. Both are not valid for highly
transparent systems in a resistive environment like perpendicular metallic
spin valves. The solution to this problem is a renormalization of the
conventional and spin-mixing conductance parameters.

\end{abstract}

\section{Introduction}

The giant magnetoresistance, as well as most of the current
magnetoelectronics, can be understood in terms of the transport of electrons
in either spin-up or spin-down state, since the magnetizations are collinear
(parallel or antiparallel) with the spin-quantization axis. Both charge and
spin transport can be described in terms of two \textquotedblleft
channels\textquotedblright\ with spin-dependent conductivities, scattering
rates \textit{etc}. \cite{Maekawa}. This \textquotedblleft
digital\textquotedblright\ magnetoelectronics does not profit from the
\textquotedblleft analogue\textquotedblright\ freedom of a magnetization to
point in any direction. Early seminal contributions by Slonczewski
\cite{Slon96} and Berger \cite{Berger96} revealed fundamentally new physics
and technological possibilities of non-collinearity, which triggered a large
number of experimental and theoretical studies. An important example is the
non-equilibrium spin-current induced torque (briefly, spin torque) which one
ferromagnet can exert on the magnetization vector of a second magnet through a
normal metal in a biased spin valve structure. This torque can be large enough
to dynamically turn magnetizations \cite{Tsoi98}, which is potentially
interesting as a low-power switching mechanism for magnetic random access
memories \cite{Inomata01}. Non-collinear magnetizations are also essential for
novel magnetic devices like the spin-flip \cite{Brataas00,Xia02} and
spin-torque transistors \cite{Bauer03}, detection of spin-precession
\cite{Huertas00}, the Gilbert damping of the magnetization dynamics in thin
magnetic films \cite{Tserkovnyak:prl02}, and spin-injection induced by
ferromagnetic resonance \cite{Brataas02}.

We are interested in heterostructures containing band ferromagnets that are
accurately described by a Stoner spin-density functional model. Elemental
metals and its alloys have high electron densities and their thin-film
heterostructures are usually considerably disordered. Size quantization
effects on transport can therefore mostly be disregarded \cite{Suzuki02}.
Semiclassical methods are appropriate in slowly varying bulk regions of the
structures, but heterointerfaces can be atomically sharp and must be treated
fully quantum mechanically. There are basically two methods which are suitable
to understand and compute transport properties of these systems from first
principles, \textit{viz}. Green function theory with configurational averaging
\cite{Rammer} and the scattering formalism for transport, combined with random
matrix theory \cite{Beenakker}. These two approaches have recently been
extended to non-collinear magnetic structures, \textit{viz}. magnetoelectronic
circuit \cite{Brataas00} and random matrix theory \cite{Waintal00}. Here, as
in \cite{Bauer02}, we show that both approaches are closely related, but do
not hold for transparent interfaces. Following Schep's \cite{Schep97} strategy
for collinear systems, both theories can be generalized, leading to analytical
results for perpendicular spin valves with parameters that can be obtained by
\textit{ab initio} band structure calculations, as well as determined
quantitatively by experiments.

\section{Boltzmann and Diffusion Equation}

When a local non-equilibrium magnetization does not point in the direction of
the spin-quantization axis, the distribution function for band states at the
Fermi energy with index $n$ is a matrix in Pauli spin space
\begin{equation}
\hat{f}_{n}\left(  \vec{r}\right)  =\left(
\begin{array}
[c]{cc}%
f_{\uparrow\uparrow}\left(  \vec{r}\right)  & f_{\uparrow\downarrow}\left(
\vec{r}\right) \\
f_{\uparrow\downarrow}\left(  \vec{r}\right)  & f_{\downarrow\downarrow
}\left(  \vec{r}\right)
\end{array}
\right)  _{n}=f_{n}^{c}\left(  \vec{r}\right)  \hat{1}+\text{\boldmath$\hat
{\sigma}\cdot$}\vec{f}_{n}^{s}\left(  \vec{r}\right)  .
\end{equation}
On the right hand side the distribution is expanded into unit matrix and the
vector of the Pauli spin matrices. $f_{n}^{c}$ is charge accumulation and the
spin accumulation $\vec{f}_{n}^{s}$ is a vector whose direction is always
parallel to the magnetization vector $\vec{m}$ in the bulk of a ferromagnet,
but arbitrary in a normal metal depending on device configuration and applied
biases. $\hat{f}_{n}$ can be diagonalized by unitary rotation matrices in spin
space, characterized by the polar angles $\theta$ and $\varphi.$ Let us assume
for simplicity that these angles are piecewise constant in position space,
thus disregarding magnetic domain walls \cite{Tatara} and magnetic
field-induced\ spin precession in normal metals \cite{Huertas00}. In the local
spin quantization frame the distribution function is diagonal with two spin
components $s=\pm1$. Introducing spin-conserving and spin-flip scattering life
times $\tau_{s}$ and $\tau_{s,-s}^{\mathrm{sf}},$ for the sake of argument
taken to be state-independent, and separating the distribution into an
isotropic electrochemical potential $\mu_{s}$ and an anisotropic term
$\gamma_{ns}$ that vanishes when averaged over the Fermi surface, the
Boltzmann equation for the stationary state reads \cite{VF}
\begin{equation}
\vec{v}_{is}\cdot\vec{\nabla}\left(  \gamma_{ns}+\mu_{s}\right)  +\left(
\frac{1}{\tau_{s}}+\frac{1}{\tau_{s,-s}^{\mathrm{sf}}}\right)  \gamma
_{ns}=\frac{\mu_{-s}-\mu_{s}}{\tau_{s,-s}^{\mathrm{sf}}}.
\end{equation}
where $\vec{v}_{ns}$ is the group velocity of state $ns$. Charge and spin
currents read
\begin{align}
j_{c}  &  =\frac{e}{hA}\sum_{ns}\vec{v}_{ns}\gamma_{ns}\;,\\
\vec{j}_{s}  &  =\frac{1}{4\pi A}\sum_{ns}\vec{v}_{ns}s\gamma_{ns}\;.
\end{align}
The Boltzmann equation is still unnecessarily complicated for most realistic
systems. In the presence of sufficient disorder, only the lowest harmonics of
$\gamma_{ns}$ in reciprocal space survive. In that limit, the Boltzmann
equation reduces to the diffusion equation
\begin{equation}
\nabla^{2}\left[  \mu_{s}(\vec{r})-\mu_{-s}(\vec{r})\right]  =\frac{\mu
_{s}(\vec{r})-\mu_{-s}(\vec{r})}{\ell_{sd}^{2}}\;.
\end{equation}
$\ell_{sd}=\sqrt{D\tau^{\mathrm{sf}}}$ is the spin-flip diffusion length,
which does not depend on spin index \cite{Filip}. The spin-averaged diffusion
coefficient $D$ can be written in terms of the density of states at the Fermi
energy $N_{s}\left(  E_{F}\right)  $\
\begin{equation}
\frac{1}{\left(  N_{\uparrow}\left(  E_{F}\right)  +N_{\downarrow}\left(
E_{F}\right)  \right)  D}=\frac{1}{N_{\uparrow}\left(  E_{F}\right)
D_{\uparrow}}+\frac{1}{N_{\downarrow}\left(  E_{F}\right)  D_{\downarrow}}\;.
\end{equation}
in terms of the spin-dependent diffusion coefficients. In a simple two-band
model $D_{s}=v_{s}\tau_{s}/3$, where $v_{s}$ are the spin-dependent Fermi
velocities. The average spin-flip relaxation time is defined as
\begin{equation}
\frac{1}{\tau^{\mathrm{sf}}}=\frac{1}{\tau_{\uparrow,\downarrow}^{\mathrm{sf}%
}}+\frac{1}{\tau_{\downarrow,\uparrow}^{\mathrm{sf}}}\;.
\end{equation}
The currents
\begin{equation}
j_{s}(\vec{r})=-\frac{\sigma_{s}}{e}\vec{\nabla}\mu_{s}(\vec{r})
\end{equation}
are governed by the spin-dependent conductivities%
\begin{equation}
\sigma_{s}=N_{s}\left(  E_{F}\right)  e^{2}D_{s}\;.
\end{equation}

\section{Boundary Conditions}

The semiclassical approach is valid when the potential landscape varies slowly
on the scale of the Fermi wave length. In heterostructures we often encounter
regions in which materials change on an atomic scale, such as at intermetallic
interfaces or tunnel junctions, which have to be treated quantum mechanically.
The \textquotedblleft nodes\textquotedblright\ are the bulk regions, in which
the semiclassical distributions are well defined. The intermediate
scattering\ regions, or \textquotedblleft contacts\textquotedblright, can then
be treated formally exactly by boundary conditions, which link the
distributions of two neighboring nodes.
\begin{figure}
[ptb]
\begin{center}
\includegraphics[
height=2.7326cm,
width=10.177cm
]%
{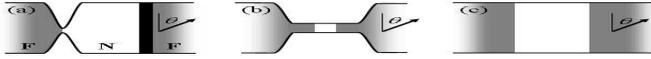}%
\caption{Different realizations of perpendicular spin valves in which $\theta$
is the angle between magnetization directions. (a) Highly resistive junctions
like point contacts and tunneling barriers limit the conductance. (b) Spin
valve in a geometrical constriction amenable to the scattering theory of
transport. (c) Magnetic multilayers with transparent interfaces}%
\end{center}
\end{figure}
Consider the spin valve structures in Fig. 1, which may be part of a larger
circuit. We denote the distribution functions in the ferromagnetic terminals
by subscripts $L$ end $R$. We explicitly allow for a drifting distribution by
the superscript $\alpha=\pm1,$ which indicates whether drift is in $\left(
\alpha=1\right)  $ or against $\left(  \alpha=-1\right)  $ the transport
direction (from left to right). Taking into account the difference between the
left- and right-moving distribution functions is the key generalization of the
previous theories in our treatment here. In Sec. 5 it is discussed how the
circuit theory can be recovered be renormalizing the conductance parameters.
Here we concentrate on scattering theory.

In order to work with simple matrices instead of diadics, we follow Waintal
\textit{et al}. \cite{Waintal00} and introduce the $4\times1$ vector
representation $\left[  \vec{f}_{n}^{\alpha}\left(  \vec{r}\right)  \right]
^{T}=\left(  f_{\uparrow\uparrow}\left(  \vec{r}\right)  ,f_{\uparrow
\downarrow}\left(  \vec{r}\right)  ,f_{\uparrow\downarrow}\left(  \vec
{r}\right)  ,f_{\downarrow\downarrow}\left(  \vec{r}\right)  \right)
_{n}^{\alpha}$. The boundary conditions for the non-equilibrium distributions
to the left and right of the scattering region then read:
\begin{align}
\vec{f}_{R,n}^{+}  &  =\sum_{m\epsilon L}\left(  \check{T}_{L\rightarrow
R}\right)  _{nm}\vec{f}_{L,m}^{+}+\sum_{m\epsilon R}\left(  \check
{R}_{R\rightarrow R}\right)  _{nm}\vec{f}_{R,m}^{-}\;,\\
\vec{f}_{L,n}^{-}  &  =\sum_{m\epsilon L}\left(  \check{R}_{L\rightarrow
L}\right)  _{nm}\vec{f}_{L,m}^{+}+\sum_{j\epsilon R}\left(  \check
{T}_{R\rightarrow L}\right)  _{nm}\vec{f}_{R,m}^{-}\;.
\end{align}
$\check{T}$, $\check{R}\ $are $4\times4$ transmission and reflection
probability matrices and the subscripts indicate the direction of the currents
($L\rightarrow R$ denotes transmission from left to right, $R\rightarrow R$
reflection from the right, \textit{etc}.). All matrix elements follow from the
scattering matrix and are conveniently normalized, for example%
\begin{equation}
\left[  \left(  \check{T}_{L\rightarrow R}\right)  _{nm}\right]  _{SS^{\prime
}}=\frac{1}{N_{S}^{L}}\left(  \vec{t}_{nm}^{L\rightarrow R}\right)
_{S}\left(  \vec{t}_{nm}^{L\rightarrow R}\right)  _{S^{\prime}}^{\dagger}.
\end{equation}
The transmission amplitudes, such as $t_{ns,ms^{\prime}}^{L\rightarrow R}$ of
a wave coming in from the left as mode $m$ and spin $s^{\prime}$ and going out
in mode $n$ and spin $s,$ are here collected in vectors $\vec{t}%
_{nm}^{L\rightarrow R}=\left(  t_{\uparrow\uparrow}^{L\rightarrow
R},t_{\uparrow\downarrow}^{L\rightarrow R},t_{\uparrow\downarrow
}^{L\rightarrow R},t_{\downarrow\downarrow}^{L\rightarrow R}\right)  _{nm}%
^{T}$. We also have $S\in\left[  1,4\right]  $, $N_{S}^{L}=N_{\uparrow}%
^{F}\left(  \delta_{S,1}+\delta_{S,2}\right)  +N_{\downarrow}^{F}\left(
\delta_{S,3}+\delta_{S,4}\right)  ,$ where $N_{s}^{F}$ is the number of modes
for spin $s$ in the ferromagnet.

In a nutshell, this is a very general formulation of charge and spin
transport, but it is not yet amenable for analytic treatment or analysis of
experiments. The isotropy assumption that reduced the Boltzmann to the
diffusion equation in the previous chapter, enormously simplifies the results,
as demonstrated in the following.

We focus here on the electrical charge current as a function of the
magnetization configuration in symmetric spin valves, as in Fig. 1(b),(c), in
order to keep the analytical manipulations manageable. We will see later that
we can derive rules from these results that are valid for general structures.
$\hat{T}$ and $\hat{R}$ are functions of the magnetic configuration, which,
disregarding magnetic anisotropies, can be parameterized by a single polar
angle $\theta.$ In first instance, we disregard spin-flip scattering and
discuss later how it can be included. Integrating over the lateral coordinates
leaves a position dependence only in the transport direction ($x$). The next
step is the assumption that the distribution functions for incident electrons
from the left and right are isotropic in space. The distribution functions for
the outgoing electrons do not have to be isotropic, as long as they are
subsequently scrambled in the nodes. The isotropy assumption may be invoked
when the nodes are diffuse or chaotic, such that electrons are distributed
equally over all states at the (spin-dependent) Fermi surfaces (which is
equivalent to replacing state dependent scattering matrix elements by its
average \cite{Schep97}). The Fermi surface integration is then carried out
easily, and the distribution functions within left and right ferromagnet
nodes\ (at locations $x_{L}$ and $x_{R}$, respectively) are matched
\textit{via} simplified boundary conditions
\begin{subequations}
\label{bound}%
\begin{align}
\vec{f}^{+}\left(  x_{R}\right)   &  =\check{T}_{L\rightarrow R}\left(
\theta\right)  \vec{f}^{+}\left(  x_{L}\right)  +\check{R}_{R\rightarrow
R}\left(  \theta\right)  \vec{f}^{-}\left(  x_{R}\right)  ,\\
\vec{f}^{-}\left(  x_{L}\right)   &  =\check{R}_{L\rightarrow L}\left(
\theta\right)  \vec{f}^{+}\left(  x_{L}\right)  +\check{T}_{R\rightarrow
L}\left(  \theta\right)  \vec{f}^{-}\left(  x_{R}\right)  ,
\end{align}
where the $4\times4$ transmission and reflection probability matrices have
elements like \cite{Waintal00}:
\end{subequations}
\begin{equation}
\left[  \check{T}_{L\rightarrow R}\right]  _{SS^{\prime}}=\frac{1}{N_{S}^{F}%
}\sum_{mn}\left(  \vec{t}_{nm}^{R\rightarrow L}\right)  _{S}\left(  \vec
{t}_{nm}^{R\rightarrow L}\right)  _{S^{\prime}}^{\dagger}.
\end{equation}
In the coordinate systems defined by the magnetization directions, the
transverse components of the spin accumulation in the ferromagnets vanish
identically \cite{Brataas00,Zangwill} and the distributions in the magnets
depend on the local spin current densities $\gamma_{s}$ and chemical
potentials $\mu_{s}$ only
\begin{equation}
\vec{f}^{\pm}\left(  x_{L/R}\right)  =\left(  \left(  \pm\gamma_{\uparrow}%
+\mu_{\uparrow}\right)  \left(  x_{L/R}\right)  ,0,0,\left(  \pm
\gamma_{\downarrow}+\mu_{\downarrow}\right)  \left(  x_{L/R}\right)  \right)
\;. \label{fF}%
\end{equation}
By this choice the explicit angle-dependence of transport is contained only in
the matrices.

We can now link an arbitrary distribution on the left to compute the
distributions on the right, subject to the constraint of charge current
conservation. Here we focus on the simple case in which we apply a bias
\begin{equation}
\Delta\mu=\sum_{s}\left(  \mu_{s}\left(  x_{L}\right)  -\mu_{s}\left(
x_{R}\right)  \right)  \;,
\end{equation}
but no spin accumulation gradient $\mu_{s}\left(  x_{L}\right)  -\mu
_{-s}\left(  x_{L}\right)  =\mu_{s}\left(  x_{R}\right)  -\mu_{-s}\left(
x_{R}\right)  $ over the system. We then find that $\gamma_{s}\left(
x_{L}\right)  =\gamma_{s}\left(  x_{R}\right)  ,$ \textit{i.e}. the spin
current component parallel to the magnetization on left and right ferromagnets
are the same. The charge current
\begin{equation}
I_{c}=\frac{e^{2}}{h}\sum_{s}N_{s}^{F}\gamma_{s}%
\end{equation}
divided by the chemical potential drop is the electrical conductance
$G=I_{c}/\Delta\mu$. Eqs. (\ref{bound},\ref{fF}) then lead to
\begin{equation}
G=\frac{2e^{2}}{h}\sum_{\substack{S=1,4\\S^{\prime}=1,4}}\left\{  N_{S}%
^{F}\left[  \check{1}-\check{T}_{L\rightarrow R}+\check{R}_{R\rightarrow
R}\right]  ^{-1}\check{T}_{L\rightarrow R}\right\}  _{SS^{\prime}}\,.
\label{GSV}%
\end{equation}
When the transparency is small, all transmission probabilities are close to
zero, reflection probabilities are close to unity, and the
Landauer-B\"{u}ttiker conductance, starting point of \cite{Waintal00}, is
recovered:%
\begin{equation}
G\rightarrow\frac{e^{2}}{h}\sum_{\substack{S=1,4\\S^{\prime}=1,4}}\left\{
N_{S}^{F}\hat{T}\left(  \theta\right)  \right\}  _{SS^{\prime}}\;. \label{LB}%
\end{equation}
Indeed, in this limit the distributions to the left and right are not
perturbed by the current, the nodes are genuine reservoirs, and standard
scattering theory applies. Also, when $\theta=0,\pi,$ Eq. (\ref{GSV}) is
equivalent to results by Schep \textit{et al}. \cite{Schep97} for the
two-channel model.

The scattering region is still not specified and may be interacting and/or
quantum coherent. We now discuss how analytical results can be obtained in the
non-interacting, diffuse limit.

\section{Semiclassical Concatenation}

The scattering matrix of a composite system can be formulated as
concatenations of the scattering matrix from separate elements, \textit{e.g..}
the scattering matrices of bulk layers and interfaces \cite{Datta}. By
assuming isotropy, \textit{i.e}. sufficient disorder or chaotic scattering,
Waintal \textit{et al}. \cite{Waintal00} proved by averaging over random
scattering matrices that size quantization effects like the equilibrium
exchange coupling or other phase coherent phenomena are destroyed by disorder
and vanish like the inverse of the number of modes. Under these conditions we
are free to define nodes in the interior of the device and link them via the
boundary conditions (\ref{bound}). This is equivalent to composing the total
transport probability matrices in Eq. (\ref{GSV}) in terms of those of
individual elements by semiclassical concatenation rules \cite{Shapiro}. The
$4\times4$ transmission probability matrix through a F$\left(  0\right)
$/N/F$\left(  \theta\right)  $ double heterojunction as in Fig.~1 in which
bulk scattering is absent, takes the form
\begin{equation}
\check{T}\left(  \theta\right)  \equiv\check{T}_{\mathrm{N\rightarrow F}%
}\left(  \theta\right)  \left[  \check{1}-\check{R}_{\mathrm{N\rightarrow N}%
}\left(  0\right)  \check{R}_{\mathrm{N\rightarrow N}}\left(  \theta\right)
\right]  ^{-1}\check{T}_{\mathrm{F\rightarrow N}}\left(  0\right)  ,
\label{concat}%
\end{equation}
where the interface transmission and reflection matrices as a function of
magnetization angle appear. The transformations needed to obtain $\check
{T}_{\mathrm{N\rightarrow F}}\left(  \theta\right)  $ and $\check
{R}_{\mathrm{N\rightarrow N}}\left(  \theta\right)  $ require some attention.
In terms of the spin-rotation
\begin{equation}
\hat{U}=\left(
\begin{array}
[c]{cc}%
\cos\theta/2 & -\sin\theta/2\\
\sin\theta/2 & \cos\theta/2
\end{array}
\right)
\end{equation}
and projection matrices $\left(  s=\pm1\right)  $%
\begin{equation}
\hat{u}_{s}\left(  \theta\right)  =\frac{1}{2}\left(
\begin{array}
[c]{cc}%
1+s\cos\theta & s\sin\theta\\
s\sin\theta & 1-s\cos\theta
\end{array}
\right)  \;, \label{us}%
\end{equation}
the interface scattering coefficients (omitting the mode indices for
simplicity) are transformed as follows \cite{Brataas00}
\begin{align}
\hat{r}_{N\rightarrow N}  &  =\hat{U}\hat{r}_{cN}\hat{U}^{\dagger}=\sum
_{s}\hat{u}_{s}r_{s}^{cN}\;,\\
t_{ss^{\prime}}^{F\rightarrow N}  &  =U_{ss^{\prime}}t_{s^{\prime}}^{cF}\;,\\
t_{ss^{\prime}}^{N\rightarrow F}  &  =t_{s}^{cN}U_{ss^{\prime}}^{\dagger}\;,\\
r_{ss^{\prime}}^{F\rightarrow F}  &  =r_{s}^{cF}\delta_{ss^{\prime}}\;.
\end{align}
The superscript $c$ indicates that the matrices should be evaluated in the
reference frame of the local magnetization, and are thus diagonal in the
absence of spin-flip relaxation scattering at the interfaces. Different
transformation properties for the different elements of the scattering matrix
derive from our choice to use local spin-coordinate systems that may differ
for each magnet. Let us, for example, inspect a transmission matrix element
from the normal metal into the ferromagnet with magnetization rotated by
$\theta$%
\begin{equation}
\left[  \check{T}_{N\rightarrow F}\left(  \theta\right)  \right]  _{11}%
=\frac{1}{N_{\uparrow}^{F}}\sum_{mn}t_{n\uparrow m\uparrow}^{R\rightarrow
L}\left(  t_{n\uparrow m\uparrow}^{R\rightarrow L}\right)  ^{\dagger}=\frac
{1}{2}\left(  1+\cos\theta\right)  \frac{1}{N_{\uparrow}^{F}}\sum
_{mn}\left\vert t_{n\uparrow m\uparrow}^{cN}\right\vert ^{2}%
\end{equation}
and analogously for the other matrix elements as well as other matrices.

Transport through a more complex system can be treated by repeated
concatenation of two scattering elements in terms of reflection and
transmission matrices analogous to Eq. (\ref{concat}). In the presence of
significant bulk scattering, we can represent a disordered metal $B$ with
thickness $d_{B}$ by diagonal matrices like \cite{Schep97,Waintal00}%
\begin{equation}
\left(  \check{T}_{B}\right)  _{SS^{\prime}}=\left(  1+\frac{1}{N_{s}^{B}%
}+\frac{e^{2}}{h}\frac{\rho_{s}^{B}d_{B}}{A_{B}}\right)  ^{-1}\delta
_{SS^{\prime}}\;,
\end{equation}
where $\rho_{s}^{B},$ $A_{B}$ are the single-spin bulk resistivities and cross
section of the bulk metal (normal or magnetic).

The \emph{interface} parameters of the present theory are the spin-dependent
Landauer-B\"{u}ttiker conductances
\begin{equation}
g_{s}=\sum_{lm}\left\vert t_{lm,s}^{cN}\right\vert ^{2}=N_{N}-\sum
_{lm}\left\vert r_{lm,s}^{cN}\right\vert ^{2}=\sum_{lm}\left\vert
t_{lm,s}^{cF}\right\vert ^{2}=N_{S}^{F}-\sum_{lm}\left\vert r_{lm,s}%
^{cF}\right\vert ^{2}%
\end{equation}
and the real and imaginary part of the spin-mixing conductance%
\[
g_{s-s}=N_{N}-\sum_{lm}r_{lm,s}^{cN}\left(  r_{lm,-s}^{cN}\right)  ^{\ast}\;,
\]
which can also be represented in terms of the total conductance $g=g_{\uparrow
}+g_{\downarrow},$ polarization $p=\left(  g_{\uparrow}-g_{\downarrow}\right)
/g,$ and relative mixing conductance $\eta=2g_{\downarrow\uparrow}/g.$

The actual concatenation of the $4\times4$ matrices defined here is rather
complicated even when using symbolic programming routines. This explains why
in \cite{Waintal00} analytic results were obtained only in special limiting
cases. We found that final results are simple even in the most general cases,
not only for Eq. (\ref{LB}) considered by \cite{Waintal00}, but also for Eq.
(\ref{GSV}). For the spin valves in Fig. 1, we find for the conductance as a
function of angle
\begin{equation}
G\left(  \theta\right)  =\frac{\tilde{g}}{2}\left(  1-\frac{\tilde{p}^{2}%
}{1+\frac{\left\vert \tilde{\eta}\right\vert ^{2}}{\operatorname{Re}%
\tilde{\eta}}\frac{1+\cos\theta}{1-\cos\theta}}\right)  \;,\label{gnew}%
\end{equation}
where $\tilde{\eta}=2\tilde{g}_{\uparrow\downarrow}/\tilde{g}\ $and
\begin{align}
\frac{1}{\tilde{g}_{s}} &  =\frac{1}{g_{s}}+\frac{e^{2}}{h}\frac{\rho
_{F,s}d_{F}}{2A_{F}}+\frac{e^{2}}{h}\frac{\rho_{N}d_{N}}{2A_{F}}-\frac{1}%
{2}\left(  \frac{1}{N_{s}^{F}}+\frac{1}{N_{N}}\right)  \label{Schep}\\
\frac{1}{\tilde{g}_{\uparrow\downarrow}} &  =\frac{1}{g_{\uparrow\downarrow}%
}+\frac{e^{2}}{h}\frac{\rho_{N}d_{N}}{2A_{N}}-\frac{1}{2N_{N}}.\label{rmix}%
\end{align}
Equation~(\ref{gnew}) is identical to the angular magnetoresistance derived by
circuit theory \cite{Brataas00} after replacement of $\tilde{g}_{s}$ and
$\tilde{g}_{\uparrow\downarrow}$ by $g_{s}$ and $g_{\uparrow\downarrow}.$
Physically, in Eqs.~(\ref{Schep},\ref{rmix}) spurious Sharvin resistances are
substracted from the interface resistances obtained by scattering theory,
whereas bulk resistances are added. These corrections are large for
transparent interfaces and essential to obtain agreement between experimental
results of transport experiments in CPP (current perpendicular to plane)
multilayers \cite{Pratt1,Gijs,GMRREV} and first-principles calculations, for
conventional \cite{Schep97,Stiles00,Xia01} as well as mixing conductances
\cite{Xia02}. The mixing conductance parameterizes the magnetization torque
due to a spin accumulation in the normal metal, governed by the reflection of
electrons from the normal metal. It is therefore natural that the mixing
conductance is reduced by the bulk resistance of the normal metal and we can
also understand that only the normal metal Sharvin resistance has to be
substracted. The real part of the mixing conductances is often close to the
number of modes in the normal metal $g_{\uparrow\downarrow}\approx N_{N}$, in
which case $\tilde{g}_{\uparrow\downarrow}\approx N_{N}/2$
\cite{Tserkovnyak02}. By letting $N_{s}^{F}\rightarrow\infty$ we are in the
regime of \cite{Waintal00}. The circuit theory is recovered when,
additionally, $N_{N}\rightarrow\infty.$ The bare mixing conductance is bounded
not only from below $\mathrm{Re}g_{\uparrow\downarrow}\geqslant g/2$%
\ \cite{Brataas00}, but also from above $\left\vert g_{\uparrow\downarrow
}\right\vert ^{2}/\mathrm{Re}g_{\uparrow\downarrow}\leqslant2N_{N}.$

\section{Extended Circuit Theory}

It is not obvious how these results should be generalized to more complicated
circuits and devices as well as to the presence of spin-flip scattering in the
normal metal. The magnetoelectronic circuit theory \cite{Brataas00} does not
suffer from these drawbacks. Originally, it was assumed in Ref.
\cite{Brataas00} that local spin and charge currents through the contacts only
depend on the generalized potential differences, and the local node chemical
potentials are obtained by a spin-generalization of the Kirchhoff laws of
electrical circuits. This is valid only for highly resistive contacts, such
that the in and outgoing currents do not significantly disturb the
quasi-equilibrium distribution of the nodes. Fortunately we are able to relax
this limitation and take into account a drift term in the nodes as well. In
order to demonstrate this, we construct the fictitious circuit depicted in
Fig.~2.%
\begin{figure}
[ptb]
\begin{center}
\includegraphics[
natheight=12.399900cm,
natwidth=22.199100cm,
height=5.9111cm,
width=10.5438cm
]%
{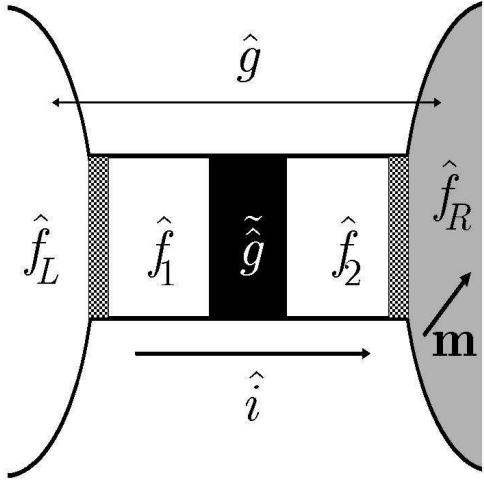}%
\caption{Fictitious device that illustrates the generalization of circuit
theory to transparent resistive elements as discussed in the text}%
\end{center}
\end{figure}
Consider a junction that in conventional circuit theory is characterized by a
matrix conductance $\hat{g}$, leading to a matrix current $\hat{\imath}$ when
the normal and ferromagnetic distributions $\hat{f}_{L}$ and $\hat{f}_{R}$ are
not equal. When the distributions of the nodes are isotropic, we know from
circuit theory that%
\begin{equation}
\hat{\imath}=\sum_{ss^{\prime}}\left(  \hat{g}\right)  _{ss^{\prime}}\hat
{u}_{s}\left(  \hat{f}_{L}-\hat{f}_{R}\right)  \hat{u}_{s^{\prime}}\,,
\end{equation}
where the projection matrices $\hat{u}_{s}$ are defined in Eq.~(\ref{us}) and
{$(\hat{g})_{ss}=g_{s},$ $(\hat{g})_{s,-s}=$}$g_{s,-s}$. Introducing lead
conductances, which modify the distributions $\hat{f}_{L}\rightarrow\hat
{f}_{1}$ and $\hat{f}_{2}\leftarrow\hat{f}_{R},$ respectively, we may define a
(renormalized) conductance matrix
$\hat{\tilde{g}}$%
, which causes an identical current $\hat{\imath}$ for the reduced (matrix)
potential drop:%
\begin{equation}
\hat{\imath}=\sum_{ss^{\prime}}\left(
\hat{\tilde{g}}%
\right)  _{ss^{\prime}}\hat{u}_{s}\left(  \hat{f}_{1}-\hat{f}_{2}\right)
\hat{u}_{s^{\prime}}\,. \label{i1}%
\end{equation}
When the lead conductances are now chosen to be twice the Sharvin
conductances, and using (matrix) current conservation
\begin{align}
\hat{\imath}  &  =2N_{N}\left(  \hat{f}_{L}-\hat{f}_{1}\right) \label{i2}\\
&  =\sum_{s}2N_{s}^{F}\hat{u}_{s}\left(  \hat{f}_{2}-\hat{f}_{R}\right)
\hat{u}_{s}\;,
\end{align}
straightforward matrix algebra leads to the result that the elements of
$\hat{\tilde{g}}$
are identical to the renormalized interface conductances found above
[Eqs.~(\ref{Schep},\ref{rmix}) without the bulk resistivities]. By replacing
$\hat{g}$ by
$\hat{\tilde{g}}$
we not only recover results for the spin valve obtained above, but we can now
use the renormalized parameters also for circuits with arbitrary complexity
and transparency of the contacts. Also spin-flip scattering in $N$ can now
also be included \cite{Brataas00}; it does not affect the form of
Eq.~(\ref{AMR}) either, but only reduces the parameter $\tilde{\chi}.$ Other
effects of the spin-flip scattering are discussed in detail by Kovalev
\textit{et al.} \cite{Kovalev02}.

\section{Applications}

Intermetallic interfaces in a diffuse environment (see Fig.~1c) have been
studied thoroughly by the Michigan State University collaboration
\cite{Pratt1} and others \cite{Gijs,GMRREV} in perpendicular (CPP) spin
valves. These experiments provided a large body of evidence for the
two-channel (\textit{i.e.} spin-up and spin-down) series resistor model and a
wealth of accurate transport parameters such as the spin-dependent interface
resistances for various material combinations
\cite{Schep97,GMRREV,Stiles00,Xia01}. In exchange-biased spin valves, it is
possible to measure the electric resistance as a function of the angle between
magnetizations, which has been analyzed experimentally and theoretically
\cite{Dauguet:prb96,Vedyayev}. Pratt \textit{c.s.} observed that experimental
magnetoresistance curves \cite{Giacomo} could accurately be fitted by the form
\cite{Brataas00}
\begin{equation}
\frac{R\left(  \theta\right)  -R\left(  0\right)  }{R\left(  \pi\right)
-R\left(  0\right)  }=\frac{1-\cos\theta}{\chi\left(  1+\cos\theta\right)  +2}
\label{AMR}%
\end{equation}
According to the new insights described above, the free parameter $\chi$ is a
function of renormalized microscopic parameters
\begin{equation}
\chi=\frac{1}{1-\tilde{p}^{2}}\frac{\left\vert \tilde{\eta}\right\vert ^{2}%
}{\mathrm{Re}\tilde{\eta}}-1
\end{equation}
in terms of the relative mixing conductance $\tilde{\eta}=2\tilde{g}%
_{\uparrow\downarrow}/\tilde{g}$, the polarization $\tilde{p}=\left(
\tilde{g}_{\uparrow}-\tilde{g}_{\downarrow}\right)  /\tilde{g}$, and the
average conductance $\tilde{g}=\tilde{g}_{\uparrow}+\tilde{g}_{\downarrow}$.

Experimental values for the parameters for Cu/Permalloy (Py) spin valves are
$\tilde{\chi}=1.2$ and $\tilde{p}=0.6$ \cite{Giacomo}. Disregarding a very
small imaginary component of the mixing conductance \cite{Xia02}, using the
known values for the bulk resistivities, the theoretical Sharvin conductance
for Cu ($0.55\cdot10^{15\;}\mathrm{\Omega}^{-1}\mathrm{m}^{-2}$/spin
\cite{Schep97}), and the spin-flip length of Py as the effective thickness of
the ferromagnet $\left(  \ell_{sd}^{F}=5\text{ nm}\ \text{\cite{Pratt1}%
}\right)  $, we arrive at the bare Cu/Py interface mixing conductance
$G_{\uparrow\downarrow}=0.39\left(  3\right)  \cdot10^{15\;}\mathrm{\Omega
}^{-1}\mathrm{m}^{-2}$. This value may be compared with the calculated mixing
conductance for a disordered Co/Cu interface ($0.55\cdot10^{15\;}%
\mathrm{\Omega}^{-1}\mathrm{m}$ \cite{Xia02}). The agreement is reasonable,
but leaves some room for material and device dependence that deserves to be
investigated in the future. The mixing conductance can also be determined from
the excess broadening of ferromagnetic resonance spectra. A larger mixing
conductance in Pt/Py can be explained by the larger density of conduction
electrons in Pt compared to Cu \cite{Tserkovnyak02}. Reasonable agreement
between experiment and theory has been also found by Zwierzycki \textit{et
al}. \cite{Zwierzycki} for Fe/Au multilayers.%

\begin{figure}
[ptb]
\begin{center}
\includegraphics[
natheight=14.036400cm,
natwidth=21.034901cm,
height=6.4383cm,
width=9.619cm
]%
{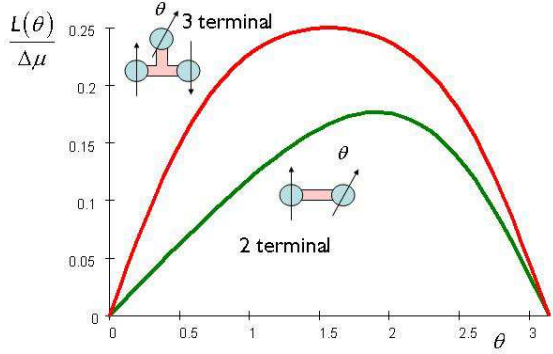}%
\caption{The spin-accumulation induced magnetization torque for a two-terminal
spin valve and a three-terminal spin-flip transistor. $\Delta\mu$ is the
source-drain bias and all contact parameters are taken to be the same, with
$\operatorname{Re}\eta=2$ and $\operatorname{Im}\eta=0$}%
\end{center}
\end{figure}
The spin torque on a ferromagnet \cite{Slon96,Waintal00} equals the spin
current through the interface with vector component normal to the
magnetization direction and its evaluation is closely related to the charge
conductance \cite{Brataas00,Waintal00}. An analytical expression for the spin
valve reads:%
\begin{equation}
L\left(  \theta\right)  =-\frac{\tilde{g}\tilde{p}\frac{\left\vert \tilde
{\eta}\right\vert ^{2}}{\operatorname{Re}\tilde{\eta}}\sin\theta}{1-\cos
\theta+\frac{\left\vert \tilde{\eta}\right\vert ^{2}}{\operatorname{Re}%
\tilde{\eta}}\left(  1+\cos\theta\right)  }\frac{\Delta\mu}{8\pi} \label{L}%
\end{equation}
Note that here the imaginary part of the mixing conductance is taken into
account explicitly, but the torque remains coplanar to the magnetization of
the contacts, \textit{i.e.} an out-of-plane \textquotedblleft
effective\textquotedblright\ field vanishes identically. Previous results
\cite{Slon96,Waintal00} are recovered in the limit that $\tilde{\eta
}\rightarrow2$ and $\tilde{p}\rightarrow1.$ By the generalized circuit theory
it is now straightforward to compute the torque on the base contact of the
spin-flip transistor with antiparallel source-drain magnetizations
\cite{Xia02}. Let us assume that the three contacts are identical, and the
base contact magnetization lies in the plane of the source and drain
magnetizations. Assuming that we may disregard spin-flip scattering in the
base contact, the in-plane torque $L_{b}$ turns out to be always larger than
the spin valve torque $L$ in the two-terminal spin valve, Eq. (\ref{L}), with
a symmetric and flatter dependence on the angle of the base magnetization
direction $\theta$ (see Fig. 3)
\begin{equation}
L_{b}\left(  \theta\right)  =-\frac{\tilde{g}\tilde{p}\operatorname{Re}%
\tilde{\eta}\sin\theta}{\left(  1-\operatorname{Re}\tilde{\eta}\right)
\cos^{2}\theta+\operatorname{Re}\tilde{\eta}+\frac{6}{2+\left\vert
\eta\right\vert ^{2}/\left(  \operatorname{Re}\tilde{\eta}\right)  ^{2}}}%
\frac{\Delta\mu}{4\pi}.
\end{equation}
In the presence of a significant imaginary part of the mixing conductance, we
also find an out-of-plane (effective field) torque $L_{\perp}\left(
\theta\right)  $ with the same angular dependence and
\begin{equation}
\frac{L_{\perp}}{L_{b}}=-2\frac{\operatorname{Im}\tilde{\eta}\operatorname{Re}%
\tilde{\eta}}{\left\vert \eta\right\vert ^{2}+2\operatorname{Re}\tilde{\eta}}.
\end{equation}

Stiles and Zangwill \cite{Stiles02} directly solved the Boltzmann equation for
spin valves to obtain angular magnetoresistance and spin torque, approximating
the mixing conductance by the number of modes (note that in a direct solution
of the Boltzmann equation this parameters should not be renormalized). The
numerical results agree well with the functional form (\ref{AMR}) (M.D.
Stiles, private communication). This function has been also derived by
Slonczewski \cite{Sloncamr} and later by Shpiro \textit{et al.} \cite{Shpiro}.
Sloncewski rederived this result with a simple circuit theory similar to that
of \cite{Brataas00} and also pointed out the relation between the angular
magnetoresistance and the spin torque. Shpiro \textit{et al.} \cite{Shpiro}
found the form (\ref{AMR}) to be valid in the limit of vanishing exchange
splitting, thus in a regime different from the transition metal ferromagnets
considered here \cite{Zangwill}.

\section{Conclusions}

We reported analytical results for the angular magnetoresistance of arbitrary
spin valves, which, by comparison with experiments \cite{Giacomo}, leads to a
value for the mixing conductance and spin torque for the Cu/Py interface of
$G_{\uparrow\downarrow}=0.39\left(  3\right)  \cdot10^{15\;}\mathrm{\Omega
}^{-1}\mathrm{m}^{-2}.$ The associated generalization of magnetoelectronic
circuit theory opens the way to engineer materials and device configurations
to optimize switching properties of magnetic random access memories. Mixing
conductances determined by experiments or first principles theory are
transferable to arbitrary devices and may be used for static as well as
dynamic transport properties. The spin-dependent interface resistances
determined by CPP-GMR\ transport experiments have played an important role in
understanding \textquotedblleft digital magnetoelectronics\textquotedblright.
We hope that the spin-mixing conductances will play a comparable role in
\textquotedblleft analogue magnetoelectronics\textquotedblright.

\textbf{Acknowledgements }We profited from discussions with Bart van
Wees, Paul Kelly, Alex Kovalev, and Yuli Nazarov, and have been supported by
FOM, NSF Grant DMR 02-33773 and the NEDO joint research program
\textquotedblleft Nano-Scale Magnetoelectronics\textquotedblright.

\end{document}